\documentclass[%
reprint,
superscriptaddress,
 amsmath,
 amssymb,
 aps,
 pra
]{revtex4-2}

\usepackage{graphicx}
\usepackage{subfig}
\usepackage{dcolumn}
\usepackage{bm}
\usepackage{caption}
\usepackage{balance}
\usepackage{bbm}
\usepackage{hyperref}
\hypersetup{
    colorlinks=true,
    linkcolor=red,
    filecolor=red,
    urlcolor=red,
    citecolor=blue, 
}

\usepackage[font=small,labelfont=bf]{caption}
\usepackage[symbol]{footmisc}
\usepackage{stackengine}

\newcommand{\sech}{\mathop{\rm sech}\nolimits}
\newcommand{\bra}[1]{\left\langle #1 \right|}
\newcommand{\ket}[1]{\left|#1\right\rangle}
\newcommand{\braket}[2]{\left\langle#1 |  #2\right\rangle}

\newcommand{\braz}[1]{\langle #1 |}
\newcommand{\ketz}[1]{|#1\rangle}

\newcommand{\lra}[2]{\langle#1\rangle}

\makeatletter
\newcommand*{\rom}[1]{\expandafter\@slowromancap\romannumeral #1@}
\makeatother

\let\f\frac
\let\p\partial

\begin{document}


\title{Describing the Wave Function Collapse Process with a State-dependent Hamiltonian}

\author{\small Le Hu}
\email{le.hu@northwestern.edu}
\affiliation{Department of Physics and Astronomy, University of Rochester, Rochester, New York 14627, USA}
\affiliation{Institute for Quantum Studies, Chapman University, 1 University Drive, Orange, CA 92866, USA}
\affiliation{Department of Physics and Astronomy, Northwestern University, Evanston, Illnois 60208, USA}
\author{\small and Andrew N. Jordan}

\affiliation{Institute for Quantum Studies, Chapman University, 1 University Drive, Orange, CA 92866, USA}
\affiliation{The Kennedy Chair in Physics, Chapman University, Orange, CA 92866, USA}
\affiliation{Department of Physics and Astronomy, University of Rochester, Rochester, New York 14627, USA}
\date{\today}

\begin{abstract}
Quantum mechanics admits two distinct evolutions: deterministic unitary dynamics governed by the Schrödinger equation and the probabilistic collapse of the wave function. We show that the continuous collapse of a quantum state under measurement can, on a trajectory-by-trajectory basis, be equivalently described as unitary evolution generated by a time- and state-dependent Hermitian Hamiltonian with stochastic parameters. While the ensemble dynamics remains non-unitary, each individual trajectory thus admits a unitary representation. We derive explicit forms of such Hamiltonians for projective measurements on arbitrary $n$-level systems and for continuous position measurements of a harmonic oscillator, and we propose experimental schemes to test these predictions. Our framework provides a new approach to modeling and controlling continuously monitored quantum systems using only state-dependent unitary resources.
 \end{abstract}

\maketitle



\section{Introduction}
Ever since the establishment of quantum mechanics a century ago, the collapse of the wave function has been its central topic much debated. Numerous interpretations and theories have been proposed, including the Copenhagen interpretation \cite{bohr1928quantum, neumann1955mathematical}, many-worlds interpretation \cite{RevModPhys.29.454}, decoherence theories \cite{adler2003decoherence}, de Broglie--Bohm theory \cite{PhysRev.85.166, PhysRev.85.180}, quantum Bayesianism \cite{PhysRevA.65.022305, fuchs2010qbism, RevModPhys.85.1693}, and many others \cite{RevModPhys.85.471}. A compelling question persists: can a theory exist that is fully compatible with the continuous framework of quantum mechanics while also accurately describing the dynamics of wave function collapse? Typically, the dynamics of a quantum state during measurement involve non-unitary evolution that depends on the measurement outcomes \cite{jordan2024quantum}. However, it is of great interest to explore whether the resulting dynamics can be reformulated solely in terms of unitary operations. This is particularly intriguing given that, during the measurement process, the atoms and molecules of both the measuring devices and the probed quantum objects are governed by the Schrödinger equation.

In this paper, we will show how the continuous collapse of the wave function can be described by the Schr\"{o}dinger equation with a time- and \textit{state}-dependent Hermitian Hamiltonian containing random parameters. The discussions are divided into pure and mixed initial quantum states, respectively. For the projective measurement on pure states, if we treat it as a fast, continuous process, then the quantum state will remain pure at any later time. The evolution history of wave function collapse can thereby be recorded as quantum trajectories $\{\ket{\psi(t_i)}\}$ \cite{jordan2024quantum}, from which we can reverse engineer and reconstruct a time-dependent Hermitian Hamiltonian $\hat{H}(t_i)$ responsible for the trajectories $\{\ket{\psi(t_i)}\}$, i.e. $\hat{H}(t_i)\ket{\psi(t_i)}=i \ket{\partial_t\psi(t_i)}$. Note that the reverse-engineering map $\ket{\psi(t)} \to \hat{H}_{\ket{\psi(t)}}$ is generally a nonlinear map, but the Hamiltonian $\hat{H}_{\ket{\psi(t)}}$ itself remains a linear operator once $\ket{\psi(t)}$ is given, different from non-linear revision of quantum mechanics e.g. \cite{bialynicki1976nonlinear, PhysRevLett.62.485,doebner1992general,grigorenko1995measurement}. Moreover, the reverse-engineered Hamiltonians $\{\hat{H}(t_i)\}$ are nonunique, and will contain random parameters $\{r_{t_i}\}$, resulting from the randomness of the trajectories $\{\ket{\psi(t_i)}\}$ originated from the interactions with the measurement device.

Therefore, even though wave function collapses are generally non-unitary processes, by focusing on event-by-event quantum trajectories, they can be reinterpreted as members of another set of unitary processes, for which we can reverse engineer the Hamiltonian for each realization of the quantum processes. To demonstrate, we solve analytically for the Hamiltonian responsible for measurements on an arbitrary $n$-level system and position measurements on an harmonic oscillator prepared in the ground state, and propose a few falsifiable experimental schemes to test this approach. Then we discuss how the above formalism developed for pure state can be generalized to the mixed states.


\section{Wave function collapse and continuous quantum measurement.}
While the projective measurement is usually treated for convenience to be happened instantaneously, it is reasonable to expect that actual physical processes always take some finite time to complete \cite{minev2019catch, mandelstam1945uncertainty, margolus1998maximum, deffner2013energy}. If we continuously measure a quantum object such that the density matrix $\hat{\rho}(t)$ of the object change gradually rather than all in a sudden, then this constitutes a continuous quantum measurement \cite{caves1987quantum, diosi1988continuous, brun2002simple, korotkov1999continuous, jacobs2006straightforward, chantasri2013action}. This process, when written in the infinitesimal form, can be described by some measurement operator $\hat{\mathcal{M}}_{\delta t}$ such that the density matrix $\hat{\rho}(t)$ at later time $t+\delta t$ changes to 
\begin{equation} \label{eq1}
\hat{\rho}(t+\delta t)=\frac{\hat{\mathcal{U}}_{\delta t} \hat{\rho}(t) \hat{\mathcal{U}}_{\delta t}^{\dagger}}{\operatorname{Tr}{[\hat{\mathcal{U}}_{\delta t} \hat{\rho}(t) \hat{\mathcal{U}}_{\delta t}^{\dagger}]}}, \quad \hat{\mathcal{U}}_{\delta t} \equiv e^{-i \hat{H}_s(t) \delta t} \hat{\mathcal{M}}_{\delta t},\end{equation}
where $\hat{H}_s(t)$ denotes the Hamiltonian of the system, $\hat{\mathcal{U}}_{\delta t}$ denotes the total evolution contributed by both $\hat{H}_s(t)$ and the measurement effect, and $\operatorname{Tr}{[\hat{\mathcal{U}}_{\delta t} \hat{\rho}(t) \hat{\mathcal{U}}_{\delta t}^{\dagger}}]$ serves as a normalization factor. Now if we consider the limit where the continuous quantum measurement disturbs the system strongly and happens extremely fast, then the whole process essentially becomes the projective measurement \cite{caves1987quantum} and can therefore be regarded as the continuous collapse of the wave function. In this extreme case, the effect from the measurement dominates such that we can take $\hat{H}_s(t) \approx 0,$ which will be assumed to be the case for the rest of the paper. Eq.~(\ref{eq1}) can thereby be rewritten as $\hat{\rho}(t+\delta t)=\frac{\hat{\mathcal{M}}_{\delta t} \hat{\rho}(t) \hat{\mathcal{M}}_{\delta t}^{\dagger}}{\operatorname{Tr}{[\hat{\mathcal{M}}_{\delta t} \hat{\rho}(t) \hat{\mathcal{M}}_{\delta t}^{\dagger}]}}$. For a given $\mathcal{\hat{M}}_{\delta t}$, we can then calculate the evolution of $\hat{\rho}(t)$ and record its evolution history as quantum trajectories \cite{Carmichael1993, wiseman1996quantum, brun2002simple, Bassi_2005, daley2014quantum}.

\section{Reconstructing the Hamiltonian from the quantum trajectories.}
Suppose we are given a history of wave function collapse, recorded as quantum trajectories $\{\hat{\rho}(t_i)\}$. For simplicity, let us first consider the case where $\hat{\rho}(0)$ is initially pure. By requiring that during the projective measurement, the quantum state does not become more mixed (since information is being extracted from the system), we can see that $\hat{\rho}(t)$ will remain pure at any later time $t$. The unchanged purity indicates that for each specific trajectory, its evolution is unitary and can therefore be described by the Schr\"{o}dinger equation, at least effectively or phenomenologically. As a result, the problem becomes that with a given history of quantum trajectory $\ket{\psi(t)}$, which could either be calculated theoretically from certain $\hat{\mathcal{M}}_{\delta t}$, or reconstructed experimentally from measurement records (e.g. via quantum tomography \cite{murch2013observing, weber2014mapping, jullien2014quantum, PhysRevX.6.041052}), how one can find an effective Hermitian Hamiltonian $\hat{H}(t)$ acting on the system alone that is responsible for the trajectory. This problem, which is equivalent to solving the Schr\"{o}dinger equation inverse problem, has been solved in our recent work \cite{hu2022quantum}. We showed that for a given continuous quantum trajectory $\ket{\psi(t)}$, one can reconstruct its corresponding Hamiltonian $\hat{H}(t)$ by
\begin{equation} \label{eq2}
	\hat{H}(t) = i (\ketz{\p_t \tilde{\psi}(t)}\braz{\tilde{\psi}({t})}-\ketz{\tilde{\psi}(t)}\braz{\p_t\tilde{\psi}(t)})+\dot{\phi}(t) \mathbbm{1},
\end{equation}
where $\phi(t)\equiv\int-i\left\langle\partial_{t} \psi(t)|\psi(t)\right\rangle d t \in \mathbb{R}$ and $\ketz{\tilde{\psi}(t)}\equiv e^{i \phi(t)}\ket{\psi(t)}$, such that $\hat{H}(t)\ket{\psi(t)}=i \ket{\p_t \psi(t)}$. Even though the reconstruction process as given by Eq.\,(\ref{eq2}) is nonlinear, this Hamiltonian itself remains linear once $\ket{\psi(t)}$ is given, such that $\hat{H}(t)(\alpha \ket{\psi(t)}+\beta{\ket{\phi(t)}})=\alpha \hat{H}(t)\ket{\psi(t)}+\beta \hat{H}(t)\ket{\phi(t)}$ for arbitrary states $\ket{\psi(t)}$ and $\ket{\phi(t)}$ and complex numbers $\alpha$ and $\beta$; see Supplemental Materials I. It also remains Hermitian, evident by directly showing $\hat{H}(t)=\hat{H}^\dagger(t)$. Hence, any evolution caused solely by this Hamiltonian will be unitary. 

The general procedures for finding the measuring Hamiltonian responsible for the continuous wave function collapse of pure states are as follows:
\begin{enumerate}
	\item Assuming the process to be time-continuous, calculate theoretically (or measure experimentally) how the quantum state $\ket{\psi(t)}$ changes infinitesimally to $\ket{\psi(t+\delta t)}$, for instance, via certain  measurement operator $\hat{\mathcal{M}}_{\delta t}$,
\begin{equation} \label{eq3}
	\ket{\psi(t+\delta t)}=\f{\hat{\mathcal{M}}_{\delta t}}{\sqrt{\bra{\psi(t)}\hat{\mathcal{M}}_{\delta t}^\dagger \hat{\mathcal{M}}_{\delta t} \ket{\psi(t)}}} \ket{\psi(t)}.
\end{equation}
where ${\sqrt{\bra{\psi(t)}\hat{\mathcal{M}}_{\delta t}^\dagger \hat{\mathcal{M}}_{\delta t} \ket{\psi(t)}}}$ serves merely as a normalization factor.
	\item Calculate $\ket{\p_t \psi(t)}$ using its limit definition (or approximate with experimental data),
	\begin{equation} \label{eq4}
		\ket{\p_t \psi(t)}=\lim_{\delta t\to0} \f{\ket{\psi(t+\delta t)}-\ket{\psi(t)}}{\delta t}.
	\end{equation}
	\item Make use of Eq.~(\ref{eq2}) to reconstruct the measuring Hamiltonian $\hat{H}(t)$.
\end{enumerate}

Note that the measuring Hamiltonian $\{\hat{H}(t_i)\}$ will contain random parameters $\{r_{t_i}\}$ coming from the readouts during the measurement, such that it is compatible with the random nature of quantum mechanics. Since for each set of measurements we will have different readouts $\{r_{t_i}\}$, the Hamiltonian will generally be different for different trajectories, i.e. it is state-dependent on the trajectories. 

Such state-dependency comes from two facts. The first is that one cannot perform exactly the same measurement twice, because of the vacuum fluctuation \cite{murch2013observing} and the enormous number of degrees of freedom in the interaction between classical measuring devices and quantum objects. In other words, every realization of measurement involves a unique microscopic configuration of the measuring device, even though macroscopically it appears not to be so. Therefore, each measurement involves a unique time-dependent measuring Hamiltonian resulting from the real-time measuring interaction. The second fact is that one is unable to know what the actual time-dependent measuring Hamiltonian is, hence the best one can do is to reverse engineer and reconstruct the measuring Hamiltonian from the measuring data (i.e. quantum trajectory) (Supplemental Materials II), the practice of which inevitably entails state-dependency. However, despite the non-uniqueness of measuring Hamiltonian, they share the same energy variance $\left\langle\partial_{t} \psi(t)|\partial_{t} \psi(t)\right\rangle-\left|\left\langle\psi(t) |\partial_{t} \psi(t)\right\rangle\right|^{2}$.

The above interpretation differs from that of objective collapse theories, which holds that the quantum mechanics is intrinsically stochastic, even for the evolution of an isolated system \cite{RevModPhys.85.471}. It also differs from that of all sorts of hidden variable theories, since our formalism assumes it is the measuring device itself that causes the random nature of quantum mechanics.

Below we will show a few examples demonstrating our formalism.

\section{Examples}
\subsection{Example I -- Two level system.}
To begin with, let us first consider the wave function of a two level system initially being at $\ket{\psi(0)}=a(0)\ket{0}+b(0)\ket{1}$, where $|a(0)|^2+|b(0)|^2=1$. We take our measuring operator $\hat{\mathcal{M}}_{\delta t}$ to be \cite{jacobs2014quantum}
\begin{equation} \label{eq5}
	\hat{\mathcal{M}}_{\delta t}=(\delta t / 2 \pi \tau)^{1 / 4} \exp \left[-\frac{\delta t}{4 \tau}\left(r_t \mathbbm{1}-\hat{\sigma}_{z}\right)^{2}\right],
\end{equation}
the physical meaning of which is that during the short time interval $[t,t+\delta t]$, we are measuring the $z$ component of the qubit with the measuring strength $1/\tau$, and the instantaneous readout $r_t \in (-\infty, +\infty)$ \cite{PhysRevLett.60.1351} at time $t$, which is a random variable, follows the conditional probability distribution $P(r_t\,|\,\rho(t))=\operatorname{Tr}[\hat{\rho}(t) \hat{\mathcal{M}}_{\delta t}\hat{\mathcal{M}}_{\delta t}^\dagger]$. The Gaussian part of the above equation indicates that the measurement effect is weighted towards the two eigenvalues $+1$ and $-1$ of the Pauli matrix $\hat{\sigma}_z$, so that the outputs of sequential applications of $\hat{\mathcal{M}}_{\delta t}$ will at last conform with the empirical facts that quantum states are always found to be at their eigenstates after von Neumann measurements. Note that different measurement approaches may give rise to different choices of $\hat{\mathcal{M}}_{\delta t}$. 
\begin{figure}[!btp]
\subfloat{\includegraphics[width=.36\textwidth,scale=1]{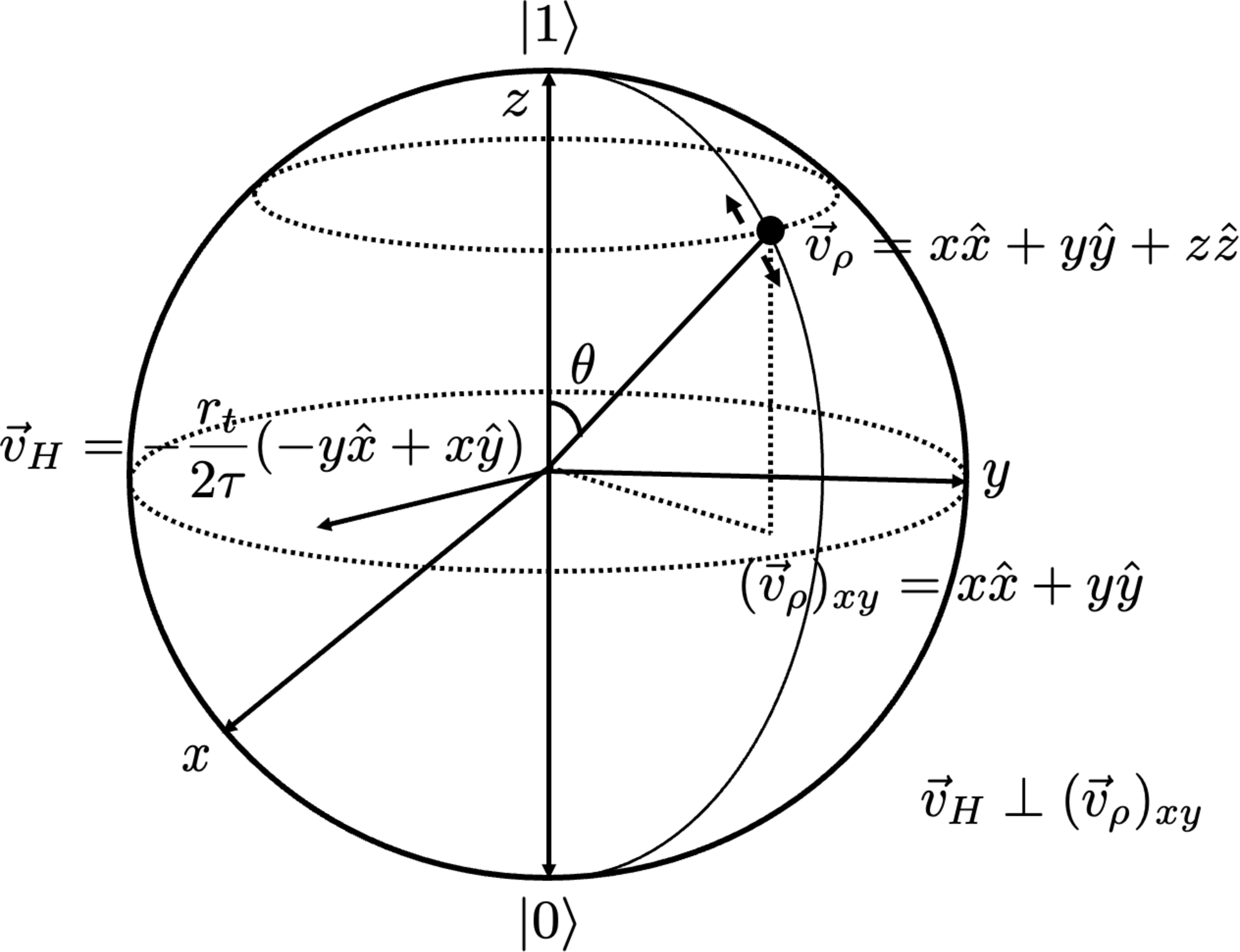}\hfill}
\caption{\label{fig:epsart1} The Bloch vector $\vec{v}_\rho$ of qubit is always orthogonal to and rotates around the Bloch vector $\vec{v}_H$ of the Hamiltonian given by Eq.\,(\ref{eq6}) during the measurement process, as visualized for two-level case. This is also the case in $n$-level system. The qubit then moves stochastically on the displayed longitudinal arc, until it stabilizes at $\ket{0}$ or $\ket{1}$. When one simultaneously measure the qubit along different axes (e.g. $\hat{x}$ and $\hat{z}$), it will make the qubit subject to the rotations of two different axes, $\vec{v}_{H_1}$ and $\vec{v}_{H_2}$, which is highly chaotic, and also makes the original eigenstates no longer be stabilization points.}
\end{figure}
Assuming the measurement is strong so that the Hamiltonian of the system itself can be ignored, one then may calculate how $\ket{\psi(0)}$ will evolve under sequential applications of $\hat{\mathcal{M}}_{\delta t}$, until it stabilizes around its eigenstate $\ket{0}$ or $\ket{1}$.

By applying the aforementioned formalism we developed [Eqs.~(\ref{eq2})-(\ref{eq4})],  we obtain the measuring Hamiltonian 
\begin{equation} \label{eq6}
\begin{aligned}
	\hat{H}(t)&=\begin{pmatrix}
	0&ia(t) b^*(t) \f{r_t}{\tau}\\
	-ia^*(t) b(t) \f{r_t}{\tau}&0
\end{pmatrix}\\
&=-\frac{r_t}{2 \tau}\left(-y(t) \hat{\sigma}_{x}+x(t) \hat{\sigma}_{y}\right),
\end{aligned}
\end{equation}
where $x(t)=2\operatorname{Re}[a(t)b^*(t)]$ and $y(t)=-2\operatorname{Im}[a(t)b^*(t)]$ are the components of the Bloch vector $\vec{v}_\rho$ of the density matrix. It can be easily verified that the $\hat{\mathcal{M}}_{\delta t}$ introduced in Eq.\,(\ref{eq5}) and the Hamiltonian we just obtained results in exactly the same evolution of $\ket{\psi(t)}$. The random variable $r_t$ is chosen from the distribution $P(r_t\,|\,\rho(t))$ depending on $a(t)$ and $b(t)$, both of which go into $H(t)$, which in turn determines the value of $a(t+\delta t)$ and $b(t+\delta t)$. In Supplemental Materials III, we show in detail the generalization of the above results in the $n$-level system.

There are a few comments worth mentioning regarding the Hamiltonian. First, it has instantaneous eigenvalues $E_{\pm}=\pm|a(t)||b(t)| \frac{r_t}{\tau}$ and instantaneous energy variance $[\Delta H(t)]^2= |a(t)|^2 |b(t)|^2 \f{r_t^2}{\tau^2}$ such that when we make a very strong measurement (i.e. $1/\tau \to \infty$), its eigenvalues and energy variance diverge (or more physically, become very large). Not only does this validate our earlier treatment $H_s(t) \approx 0$, but also it is consistent with the Heisenberg uncertainty principle $\Delta H \Delta t \gtrsim \hbar/2$, and the idea of quantum speed limit \cite{mandelstam1945uncertainty, margolus1998maximum, deffner2013energy, hu2022quantum}, that a rapidly changing quantum state inevitably has a large energy variance. The fact that $[\Delta H(t)]^2 \to \infty$ as $1/\tau \to \infty$ also echoes our earlier assumption that all physical processes take finite amount of time to happen, since in reality the energy supplied during measurement cannot be infinite. This is consistent with the example constructed by Aharonov and Bohm \cite{PhysRev.122.1649}, where they demonstrate that it is possible to measure energy with arbitrary precision within an arbitrarily short time. However, this requires that the interaction strength, which is proportional to the energy uncertainty of the interaction Hamiltonian, be adjusted to a sufficiently large value.

Second, when the quantum state reaches one of its eigenstate, $\ket{0}$ or $\ket{1}$, which means $a(t)=0$ or $b(t)=0$, the measuring Hamiltonian vanishes, so that the quantum state is stabilized and no longer evolves, had we assumed $H_s(t)\approx 0$. In addition, while the mean energy of the Hamiltonian is zero, an arbitrary real function $\dot{\phi}(t)$ can be added to the diagonal term such that $H(t) \to H(t)+\dot{\phi}(t) \mathbbm{1}$, which only changes the global $U(1)$ phase of $\ket{\psi(t)}$ and therefore can also be used to describe the measuring process. Finally, if we plot on the Bloch sphere the evolution of the qubit during the measurement, we notice that the qubit rotates exactly around the axis $-\f{r_t}{2\tau}(-y(t)\hat{x}+x(t)\hat{y})$ pointed by the Bloch vector $\vec{v}_H$ of the Hamiltonian, and that the Bloch vector $\vec{v}_\rho=x(t)\hat{x}+y(t)\hat{y}$ of the qubit is orthogonal to $\vec{v}_H$ (Fig.\,\ref{fig:epsart1}). It can be shown that this conclusion actually also applies to the measurement on an $n$-level system, if we reconstruct the Hamiltonian via Eq.\,(\ref{eq2}) (Supplemental Materials IV). Moreover, this observation illustrates why we cannot make precise simultaneous measurements in the different bases, such as $\hat{x}$ and $\hat{z}$, as this will force the qubit to stochastically rotate around ``two'' different axes, $\vec{v}_{H_1}$ and $\vec{v}_{H_2}$, making the total evolution highly chaotic. It also makes the measuring Hamiltonian no longer vanish at the eigenstates of $\hat{\sigma}_x$ and $\hat{\sigma}_z$, so the qubit will continue to evolve and escape from its original eigenstates, making the readouts spread over a large region \cite{ruskov2010qubit, chantasri2018simultaneous}.

To experimentally test the conclusions drawn from the above discussions, one can implement the following kicked quantum nondemolition measurements \cite{PhysRevB.74.085307} for a qubit system:
\begin{enumerate}
	\item Prepare a known initial state $\ket{\psi(0)}=\ket{\hat{n}}$, where $\hat{n}$ denotes an arbitrary Bloch vector. 
	\item Make a weak measurement, lasting time $\tau$, about $z$-axis, with a random readout $r_t$.
	\item Apply the counter-measuring Hamiltonian $-\hat{H}(t)$ [negative of Eq.\,(\ref{eq6})] for time $\tau$, so as to ``push back'' the qubit to where it used to be.
	\item Wait for the time period of $T-2\tau$ (assuming $T \gg \tau$), where $e^{-i H_s T}=\mathbbm{1}$. This wait voids the contribution from the system Hamiltonian.
	\item Make an projective measurement along the basis $\ket{\hat{n}}\bra{\hat{n}}-\ket{\hat{n}_\perp}\bra{\hat{n}_\perp}$, and calculate the probability that the qubit remains at $\ket{\hat{n}}$.
\end{enumerate}
In principle, if the counter-measuring Hamiltonian indeed represents the effective Hamiltonian experienced by the system, then we should have 100\% probability in detecting the qubit to be $\ket{\hat{n}}$. Otherwise our theory will be falsified. The above procedures can also be used as a novel protocol for quantum feed-back control.

Another way to experimentally test and utilize the conclusions is related to the most probable path. In \cite{PhysRevA.88.042110}, it is shown by using the stochastic path integral and the variational principle, later verified experimentally in \cite{weber2014mapping}, that there exists a most probable path of a continuously monitored quantum system, the evolution of which are subject to the effects from both the Hamiltonian of the system and the measuring processes. We can thereby solve for an effective Hamiltonian $\hat{H}_\text{eff}(t)$ responsible for such most probable path via Eq.\,(\ref{eq2}-\ref{eq4}), and by applying the counter-Hamiltonian $-\hat{H}_\text{eff}(t)$, to cancel out the most probable path. Experiments can then be designed to check whether the most probable path becomes the identity under the application of $-\hat{H}_\text{eff}(t)$ so as to test if $\hat{H}_\text{eff}(t)$ indeed describes the dynamics of a continuously monitored quantum system. In Supplemental Materials \rom{5}, we provide an example to show the calculation of $\hat{H}_\text{eff}(t)$ of a continuously monitored qubit with its Hamiltonian $\hat{H}=\f{\epsilon}{2}\hat{\sigma}_z$.

%

\subsection{Example II -- harmonic oscillator in the ground state.}
Let us consider the ground state wave function of an harmonic oscillator peaked around $x_0$ (one can take $x_0=0$ for the ground state) and at time $t=0$,
\begin{equation}
	\psi(x, 0)=\left(\frac{m \omega}{\pi \hbar}\right)^{1 / 4} \exp \left(-\frac{m \omega}{2 \hbar}\left(x-x_{0}\right)^{2}\right).
\end{equation}
The measurement operator for a weak position measurement can be described by \cite{caves1987quantum} $\hat{\mathcal{M}}_{\delta t}=\int(\delta t / 2 \pi \tau)^{1 / 4} \exp \left[-\frac{\delta t}{4 \tau}(r_t-x)^{2}\right]|x\rangle\langle x| d x$, where $1/\tau$ is the measuring strength. Expressing the wave function $\psi(x, 0)$ by the ket state $\ket{\psi(0)}=\int \psi(x,0)\ket{x}dx$, and applying Eq.\,(\ref{eq2})-(\ref{eq4}), we obtain (Supplemental Materials \rom{6}) the measuring Hamiltonian at time $t=0$,
\begin{widetext}
\begin{equation}\label{eq10}
	\hat{H}(0)=i\hbar\int \f{\eta_{x x^\prime}}{4\tau}\sqrt{\f{m \omega}{\pi \hbar}}\exp{\left( -\f{m \omega ((x-x_0)^2+{(x^\prime}-x_0)^2)}{2\hbar}\right)} \ket{x}\bra{x^\prime}dxdx^\prime,
\end{equation}
\end{widetext}
where $\eta_{xx^\prime} \equiv (x-x^\prime)(2r_t-x-x^\prime)$, which has instantaneous energy uncertainty 
\begin{equation}\label{eq11}
	\Delta H(0)=\sqrt{\frac{\hbar \left(\hbar+4 m \omega (r_t-x_0)^{2} \right)}{32 m^{2} \omega^{2}\tau^{2} }} \hbar.
\end{equation}
It can be shown that at time $t=\delta t$, the wave function is still a Gaussian, so we can apply the same procedure repeatedly, which in turn indicates that the wave function will always be a Gaussian at any later time. The corresponding measuring Hamiltonian $\hat{H}(t)$ will therefore also share a similar form as $\hat{H}(0)$, except that the $x_0$ and $\omega$ in $\hat{H}(0)$ will need to be replaced by some other numbers. More specifically, it can be shown by solving a differential equation that the position uncertainty $\Delta X(t)$ at time $t$ can be related to $\Delta X(0)$ by $[\Delta X(t)]^{2}=\frac{[\Delta X(0)]^{2} \tau}{[\Delta X(0)]^{2} t+\tau}$. By replacing $\omega$ with $\Delta X(t)$ and using Eq.\,(\ref{eq11}), we obtain
\begin{equation} \label{eq12}
	 \Delta H(t)=\frac{\sqrt{2 (r_t-x_t)^{2} [\Delta X(t)]^{2}+[\Delta X(t)]^{4}}}{2 \sqrt{2} \tau} \hbar,
	 \end{equation}
which limits to $|r_t-x_t|\hbar/(2\sqrt{rt})$ as $\tau \to 0.$
Such expression of $\Delta H(t)$ gives the energetic resources needed to quantify a given measurement dynamics. It is interesting to see that if the readout $r_t$ is too far away from the mean position $x_t$, then $\Delta H(t)$ becomes large. It is also worth noting that the Hamiltonian $\hat{H}(0)$ is non-local, since $\bra{x}\hat{H}(0)\ket{x^\prime}$ is generally nonzero for $x \neq x^\prime$. 

\section{Measurement of a mixed quantum state.}

Here we briefly discuss the generalization of the measurement on a mixed state. Consider an ensemble of (proper) mixed states $\rho(0)=\sum_i^n p_i(0) \ket{\psi_i(0)}\bra{\psi_i(0)}$ at $t=0$, which is prepared by mixing up $n$ orthonormal pure states $\ket{\psi_i(0)}$ with relative population $p_i(0)$. Then, for a particle randomly selected from the ensemble, there is a classical probability $p_i(0)$ for it to be at $\ket{\psi_i(0)}$. Now suppose one makes a measurement on the selected particle, then according to the earlier theory we developed, one concludes that there should be a Hamiltonian responsible for the measuring process, since the selected particle is actually pure --- it is just that one has lost track (information) of the state. In this case, one can describe the measuring process by telling that at time $t=0$, there is a probability $p_i(0)$ that the measuring Hamiltonian is given by $\hat{H}_{\ket{\psi_i(0)}}$. Such description $\{ p_i(0)\}$ needs to be Bayesian updated at later time $t$ to $\{ p_i(t)\}$ once a series of readouts $\{r_t\}$ are accumulated, indicating the process of purification. Note that in this scheme, the selected particle actually remains pure all the time, corresponding to a unitary process. It is the observer's updating knowledge of the state that makes the evolution appear non-unitary for the observer, i.e. probabilistic combinations of unitary evolutions can appear non-unitary \cite{hu2023probabilistic}.

For improper mixtures $\rho_A=\operatorname{Tr}_B(\rho_{AB})$, which is obtained from partial tracing over the pure state $\rho_{AB}$ living in a larger Hilbert space, the above argument also applies since improper mixtures are physically indistinguishable from proper mixtures. Alternatively, one can obtain the measuring Hamiltonian by letting the measurement operator to be $\hat{\mathcal{M}}_{\delta t} \otimes I$, and apply Eqs.\,(\ref{eq2}-\ref{eq4}) to $\rho_{AB}$, which could also account for the collapse of $\rho_A$. It is worth noting that the measuring Hamiltonian obtained this way will be non-local.

We note that once we go beyond a single quantum system and describe the quantum state dynamics of nonlocally separated systems, information must be instantaneously shared to coordinate the appropriate Hamiltonians that need to be reverse engineered.  Perhaps the most fruitful conclusion we can draw from this observation is not that we have put forward a new fundamental theory of measurement, for which indeed as we point out is physically problematic once we go beyond a single quantum system, but rather we have a new way of describing and modeling continuous quantum measurement processes, using the repeated application of unitary operators, instead of needing to apply either projectors or repeated non-unitary positive operators to the system.

\section{Conclusion}
In this paper, we demonstrate that by focusing on a single state at a time, how one can describe the wave function collapse process  by using the Schr\"{o}dinger equation and Eqs.\,(\ref{eq2})-(\ref{eq4}). We solve analytically for the measuring Hamiltonian [Eqs.\,(\ref{eq6}),(\ref{eq7}), and (\ref{eq10})] responsible for the process in varied cases, and calculate its instantaneous energy variance [Eqs.\,(\ref{eq8}) and (\ref{eq12})]. We then discuss how our results can also be applied to mixed states. Our main conclusion is that we can identify a quantum trajectory that is a member of a non-unitary process as also being a member of another, unitary process.

\section{Acknowledge} We are grateful to the support from the Army Research Office (ARO) under the grant W911NF-22-1-0258.
\section{Data Availability Statement}This work is theoretical and does not involve any data.

\appendix
\bibliographystyle{ieeetr}
\bibliography{apssamp3}
\clearpage
\newpage
 \onecolumngrid
  \begin{center}
\textbf{\large Supplemental Materials I}
\end{center}
To show that the time- and state-dependent Hermitian Hamiltonian $\hat{H}(t)$ given by Eq.\,(\ref{eq2}) is a linear operator, we only need to show that for arbitrary state vectors $\ket{a(t)}$ and $\ket{b(t)}$, and complex numbers $\alpha$ and $\beta$, we always have
\begin{equation}
	\hat{H}(t)(\alpha\ket{a(t)}+\beta \ket{b(t)})=\alpha \hat{H}(t)\ket{a(t)}+\beta \hat{H}(t) \ket{b(t)}.
\end{equation}
This is evident by
\begin{equation}
\begin{aligned}
	\hat{H}(t)(\alpha\ket{a(t)}+\beta \ket{b(t)}) &= \left[i (\ketz{\p_t \tilde{\psi}(t)}\braz{\tilde{\psi}({t})}-\ketz{\tilde{\psi}(t)}\braz{\p_t\tilde{\psi}(t)})+\dot{\phi}(t) \mathbbm{1}\right](\alpha\ket{a(t)}+\beta \ket{b(t)})\\
	&=\alpha \left(i \ketz{\p_t \tilde{\psi}(t)}\braz{\tilde{\psi}({t})}\right) \ket{a(t)} - \alpha \left(i\ketz{\tilde{\psi}(t)}\braz{\p_t\tilde{\psi}(t)}\right) \ket{a(t)}+ \alpha \dot{\phi}(t)\mathbbm{1} \ket{a(t)}\\
	&+\beta \left(i \ketz{\p_t \tilde{\psi}(t)}\braz{\tilde{\psi}({t})}\right) \ket{b(t)} - \beta \left(i\ketz{\tilde{\psi}(t)}\braz{\p_t\tilde{\psi}(t)}\right) \ket{b(t)}+ \beta \dot{\phi}(t)\mathbbm{1} \ket{b(t)}\\
	&=\alpha \hat{H}(t)\ket{a(t)}+\beta \hat{H}(t) \ket{b(t)}.
\end{aligned}
\end{equation}
Hence $H(t)$ as given by Eq.\,(\ref{eq2}) is a linear operator.
 \\\\
 \begin{center}
\textbf{\large Supplemental Materials II}
\end{center}
The measuring Hamiltonian may not be unique. Two totally different Hamiltonians with different spectra may result in exactly the same quantum evolution. For example, both $\hat{H}=-\hat{\sigma}_z$ and $\hat{H}(t)=\f{1}{3}\mathbbm{1}-\f{2}{9}\sqrt{2}\cos{(2t)}\hat{\sigma}_x+\f{2}{9}\sqrt{2}\sin{(2t)}\hat{\sigma}_y-\f{8}{9}\hat{\sigma}_z$ drive $\ket{\psi(t)}=\f{e^{it}}{\sqrt{3}}\ket{0}+\sqrt{\f{2}{3}}e^{-i t}\ket{1}$, and there is no way to tell which one is the actual Hamiltonian responsible for the driving just from (hypothetically) observing the quantum trajectory of $\ket{\psi(t)}$. This implies that the measuring Hamiltonian we obtained from Eq.~(\ref{eq2})-(\ref{eq4}) may not be the actual Hamiltonian responsible for the wave function collapse. Is there a way to find the true measuring Hamiltonian? With only the information from the quantum trajectories $\ket{\psi(t)}$, the answer is negative. Nevertheless, we prove the following theorem.
\\\\
\textit{Theorem}. \textit{Let $\ket{\psi(t)}\neq0$ be an arbitrary pure state at time $t$. Let $V$ be the linear space formed by all Hermitian operators $\hat{T}(t)$ satisfying $\hat{T}(t)\ket{\psi(t)}=0.$ If one can find at least one Hamiltonian $\hat{H}(t)$ satisfying $\hat{H}(t)\ket{\psi(t)}=i \ket{\p_t \psi(t)}$, then one can find all possible Hamiltonians $\hat{H}^\prime(t)$ satisfying $\hat{H}^\prime(t)\ket{\psi(t)}=i \ket{\p_t \psi(t)}$. In particular, $\hat{H}^\prime(t)$ can be given by $\hat{H}^\prime(t)=\hat{H}(t)+\hat{T}(t)$ where $\hat{T}(t)$ is arbitrarily chosen from $V$.}
\\\\
\textit{Proof.}  Let $\hat{H}(t)$ be a fixed Hamiltonian satisfying $\hat{H}(t)\ket{\psi(t)}=i \ket{\p_t \psi(t)}$. Let $H^\prime(t)$ be an arbitrary Hamiltonian that satisfies $\hat{H}^\prime(t)\ket{\psi(t)}=i \ket{\p_t \psi(t)}$. Since $\hat{H}(t)\ket{\psi(t)}=i \ket{\p_t \psi(t)}$, we conclude $(\hat{H}^\prime(t)-\hat{H}(t))\ket{\psi(t)}=0$. By defining $T(t)\equiv\hat{H}^\prime(t)-\hat{H}(t)$, we have $T(t) \in V$.

On the other hand, let $\hat{H}(t)$ be a fixed Hamiltonian satisfying $\hat{H}(t)\ket{\psi(t)}=i \ket{\p_t \psi(t)}$ and let $\hat{T}(t)$ be an arbitrary Hermitian operator such that $\hat{T}(t)\in V$. By definition, $\hat{T}(t)\ket{\psi(t)}=0$, therefore $(\hat{T}(t)+\hat{H}(t))\ket{\psi(t)}=i\ket{\p_t \psi(t)}$. By defining $H^\prime(t) \equiv \hat{T}(t)+\hat{H}(t)$, we see that  $\hat{H}^\prime(t)\ket{\psi(t)}=i \ket{\p_t \psi(t)}$. $\square$
\\\\
Since the unknowns in $\hat{T}$(t) are $n^2$ for $n$-dimensional Hilbert space, and the corresponding Schr\"{o}dinger equation contains $n$ equations, we have that $\dim V=n^2-n \geq 2$, where $n=2,3, 4, \dots$. In other words, there are plenty of linearly independent $\hat{T}(t)$ that satisfies $\hat{T}(t)\ket{\psi(t)}=0$ when $n$ becomes large. 

The above theorem implies that even though we may never be able to know the actual measuring Hamiltonian, we can be assured that it must be one of the possible $\hat{H}^\prime(t)$, if we are able to solve for all $\hat{T}(t)$. Moreover, all such Hamiltonians $H^\prime(t)$ should have the same energy variance equal to $\braket{\p_t \psi(t)}{\p_t \psi(t)}-|\braket{\psi(t)}{\p_t\psi(t)}|^2$.
\\\\
\begin{center}
\textbf{\large Supplemental Materials III}
\end{center}
Consider the measurement on an arbitrary $n$-level system initially at $\ket{\psi(0)}=\sum_i^n a_i(0)\ket{i}$ where $\sum_i^n|a_i(0)|^2=1$. Similar to the two-dimensional case, let the measurement operator be $\hat{\mathcal{M}}_{\delta t}=\sum_{i}^{n} M_{i}(t)|i\rangle\langle i|$ where $
	M_{i}(t)=(\delta t / 2 \pi \tau)^{1 / 4} \exp [-\frac{\delta t}{4 \tau}\left(r_t-\lambda_{i}\right)^{2}]$. The $\lambda_i$ represents the $i$-th eigenvalue corresponding to the $i$-th eigenstate of all possible outcomes. By using again Eq.\,(\ref{eq2})-(\ref{eq4}), it can be shown that the measuring Hamiltonian responsible for the wave function collapse is given by (do not confuse the imaginary number $i$ with the summation index $i$),
	\begin{equation}\label{eq7}
		[H(t)]_{i j}=i \frac{a_{i}(t) a^*_{j}(t)}{4 \tau} \sum_{k}\left|a_{k}(t)\right|^{2}\left(\eta_{i k}+\eta_{k j}\right),
	\end{equation}
	where $\eta_{i j} \equiv\left(\lambda_{i}-\lambda_{j}\right)\left(2 r_t-\lambda_{i}-\lambda_{j}\right)$. The instantaneous energy variance $[\Delta H(t)]^2$ is calculated to be
\begin{equation} \label{eq8}
	[\Delta H(t)]^2=\sum_{j}^{n}\left[\frac{\left|a_{j}(t)\right|^{2}}{16 \tau^{2}}\left(\sum_{i}^{n}\left|a_{i}(t)\right|^{2} \eta_{ij}\right)^{2}\right].
\end{equation}

Below we provide a detailed calculation regarding the above results.

Let $\ket{\psi(t)}=\sum_i^n a_i(t)\ket{i}$ be an arbitrary pure state for an $n$-level system, where $\sum_i^n |a_i(t)|^2=1$ so that it is normalized. Let $\hat{\mathcal{M}}_{\delta t}=\sum_{i}^{n} M_{i}(t)|i\rangle\langle i|$ be the measurement operator where $M_{i}(t)=(\delta t / 2 \pi \tau)^{1 / 4} \exp \left[-\frac{\delta t}{4 \tau}\left(r_t-\lambda_{i}\right)^{2}\right]$ and $\lambda_i$ is the $i$-th eigenvalue corresponding to $i$-th eigenstate off all possible states when the von Neumann measurement is finished. According to Eq.\,(\ref{eq3})(\ref{eq4}), We need to calculate 
\begin{equation}
\begin{aligned}
	\lim_{\delta t \to 0}\f{\ket{\psi(t+\delta t)}-\ket{\psi(t)}}{\delta t}&=\lim_{\delta t \to 0}\left( \f{\hat{\mathcal{M}}_{\delta t}}{\sqrt{\bra{\psi(t)}\hat{\mathcal{M}}_{\delta t}^\dagger \hat{\mathcal{M}}_{\delta t} \ket{\psi(t)}}}-1\right) \ket{\psi(t)}/{\delta t}\\&=\lim_{\delta t \to 0} \left[ \sum_{j}^{n}\left(\frac{M_{j}(t)}{\sqrt{\sum_{i}^{n}\left|a_{i}(t)\right|^{2}\left|M_{i}(t)\right|^{2}}}-1\right)|j\rangle\langle j|\right]\ket{\psi(t)}/{\delta t}.
	\end{aligned}
\end{equation}
Since 
\begin{equation}
	\lim_{\delta t \to 0}\sqrt{\sum_i^n |a_i(t)|^2|M_i(t)|^2}=\left(\f{\delta t}{2\pi \tau}\right)^{1/4}-\f{\delta t}{4 \tau}\left(\f{\delta t}{2\pi \tau}\right)^{1/4} \sum_i |a_i(t)|^2(r_t-\lambda_i)^2+\mathcal{O}(\delta t^2),
\end{equation}
\begin{equation}
	  \lim_{\delta t \to 0} M_i(t)=\left(\f{\delta t}{2\pi \tau}\right)^{1/4}-\f{\delta t}{4 \tau}\left(\f{\delta t}{2\pi \tau}\right)^{1/4} (r_t-\lambda_i)^2 +\mathcal{O}(\delta t^2),
\end{equation}
we have
\begin{equation}
	\lim_{\delta t \to 0}\ \f{M_j(t)}{\sqrt{\sum_i^n |a_i(t)|^2|M_i(t)|^2}}=1-\f{\delta t}{4\tau} \sum_{i \neq j}^n |a_i(t)|^2\eta_{ij}+\mathcal{O}(\delta t^2),
\end{equation}
where $\eta_{ij} \equiv (\lambda_i-\lambda_j)(2r_t-\lambda_i-\lambda_j)$, hence 
\begin{equation}
	\lim_{\delta t\to 0} \left( \f{\hat{\mathcal{M}}_{\delta t}}{\sqrt{\bra{\psi(t)}\hat{\mathcal{M}}_{\delta t}^\dagger \hat{\mathcal{M}}_{\delta t} \ket{\psi(t)}}}-1\right)/\delta t=-\sum_j \left(\f{1}{4\tau} \sum_{i}^n |a_i(t)|^2\eta_{ij}+\mathcal{O}(\delta t)\right) \ket{j}\bra{j},
\end{equation}
therefore
\begin{equation}
	\ket{\p_t\psi(t)}_j=-a_j(t)\left(\f{1}{4\tau} \sum_{i}^n |a_i(t)|^2\eta_{ij}\right)\ket{j}.
\end{equation}
On the other hand, since
\begin{equation}
	[\braket{\psi(t)}{\p_t\psi(t)}]_j=-|a_j(t)|^2\left(\f{1}{4\tau} \sum_{i}^n |a_i(t)|^2\eta_{ij}\right)
\end{equation}
we have
\begin{equation} \label{seq15}
	\braket{\psi(t)}{\p_t\psi(t)}=\sum_j -|a_j(t)|^2\left(\f{1}{4\tau} \sum_{i}^n |a_i(t)|^2\eta_{ij}\right)
=-\sum_{i,j} \f{1}{4\tau}|a_i(t)|^2|a_j(t)|^2 \eta_{ij}=0,
\end{equation}
where we have used the fact that $\eta_{ij}=-\eta_{ji}$ and $\eta_{ii}=0$.
Since $\braket{\psi(t)}{\p_t\psi(t)}=0$, we have $\ketz{\tilde{\psi}(t)}=\ketz{\psi(t)}$. Now that we have obtained the expression of $\ketz{\tilde{\psi}(t)}$ and $\ketz{\p_t \tilde{\psi}(t)}$, by Eq.~(2), we can obtain the measuring Hamiltonian
\begin{equation}
	[H(t)]_{ij}=i \f{a_i(t) a_j(t)^*}{4\tau} \sum_k|a_k(t)|^2 \left(\eta_{ki}+\eta_{kj}\right)
\end{equation}
and energy uncertainty
\begin{equation} \label{seq8}
	\Delta H(t)= \bra{\psi(t)}H^2(t)\ket{\psi(t)}-(\bra{\psi(t)}H(t)\ket{\psi(t)})^2=\| \ket{\p_t \psi(t)}\|=\sqrt{\sum_j^n \left[\f{|a_j(t)|^2}{16\tau^2}\left( \sum_i^n |a_i(t)|^2 \eta_{ij}\right)^2\right]}.
\end{equation}
\\\\
\begin{center}
\textbf{\large Supplemental Materials IV}
\end{center}
\setcounter{equation}{0}
\setcounter{figure}{0}
\setcounter{table}{0}
\setcounter{page}{1}
\makeatletter
\renewcommand{\theequation}{S\arabic{equation}}
\renewcommand{\thefigure}{S\arabic{figure}}
Below, we will show that for an $n$-level system, the Bloch vector representation $v_H$ of the Hamiltonian $\hat{H}(t)$
obtained via Eq.\,(\ref{eq2}) is always orthogonal to the Bloch vector  $\vec{v}_\rho$ of $\rho(t)=\ket{\psi(t)}\bra{\psi(t)}$ for arbitrary $\ket{\psi(t)}$. In addition, we show that that the effects of  $\hat{H}(t)$ on $\ket{\psi(t)}$ is making $\ket{\psi(t)}$ rotate around the Bloch vector $\vec{v}_H$ of $\hat{H}(t)$.

To start with, for an arbitrary Hermitian operator $T$, it can be written under the bases of the generators $\{\Lambda_i\}$ of SU($n$) group,
\begin{equation}
T=\lambda_0 \mathbbm{1}+\sum_{i=1}^{n^2-1} \lambda_i \Lambda_i
\end{equation}
such that we can define its generalized Bloch vector $\vec{v}_{T}=(\lambda_1~\lambda_2~\dots\lambda_{n^2-1})^T.$ The generators $\{\Lambda_i\}$, for example, can be the generalized Gell-Mann matrices. In the cases where $\lambda_0$ in unimportant or implicitly assumed, the Bloch vector $\vec{v}_T$ fully characterize the Hermitian operator $T$. 

Recall that we need to show $\vec{v}_\rho \cdot \vec{v}_H=0$. This can be done via the brute force of calculation, i.e. decomposing the measuring Hamiltonian given by Eq.\,(7) and the density matrix $\rho(t)=\ket{\psi(t)}\bra{\psi(t)}$ in terms of the bases of generalized Gell-Mann matrices \cite{bertlmann2008bloch} to obtain $\vec{v}_\rho$ and $\vec{v}_H$, and then verify that indeed it is $\vec{v}_\rho \cdot \vec{v}_H=0$. It turns out that
\begin{equation}
	\vec{v}_\rho=(\{\lambda_s^{ij}\}, \{\lambda_a^{ij}\},\{\lambda^i\})
\end{equation}
\begin{equation}
	\vec{v}_H=\f{1}{4\tau}(\{\lambda_a^{ij}\Sigma_{ij}\},\{-\lambda_s^{ij}\Sigma_{ij}\},0,0,\cdots,0)
\end{equation}
where
\begin{equation}
	\Lambda_s^{ij} \equiv |j\rangle\langle k|+| k\rangle\langle j|, \quad 1 \leqslant j<k \leqslant d
\end{equation}
\begin{equation}
	\Lambda_{a}^{j k}\equiv-\mathrm{i}|j\rangle\langle k|+\mathrm{i}| k\rangle\langle j|, \quad 1 \leqslant j<k \leqslant d
\end{equation}
\begin{equation}
	\Sigma_{ij}=\sum_{k}^{n} \lambda^{k}\left(\eta_{k i}+\eta_{k j}\right)
\end{equation}
where $\eta_{i j}\equiv(E_i-E_j)(2r-E_i-E_j)$. The $E_i$ denotes the $i$-th eigenvalue corresponding to the $i$-th eigenstate after finishing the von Neumann measurement on $\ket{\psi(t)}$. The $\lambda_s^{ij}$, $\lambda_a^{ij}$, and $\lambda^{i}$ are the coefficients of the Gell-Mann matrices $\Lambda_s^{ij}$, $\Lambda_a^{ij}$ and $\Lambda^i$ \cite{bertlmann2008bloch}, respectively, where \begin{equation}
	\Lambda^{l}=\sqrt{\frac{2}{l(l+1)}}\left(\sum_{j=1}^{l}|j\rangle\langle j|-l| l+1\rangle\langle l+1|\right), \quad 1 \leqslant l \leqslant d-1.
\end{equation}
It is straightforward to verify that $\vec{v}_\rho \cdot \vec{v}_H=0$.

Another simpler way to prove $\vec{v}_\rho \cdot \vec{v}_H=0$ is by proving $\hat{\rho}(t)$ is orthogonal to $\hat{H}(t)$. This is equivalent to proving $\operatorname{Tr} [{\hat{\rho}(t)\hat{H}(t)}]=\langle \hat{H}(t)\rangle_t=0$, which is always true as shown in Eq.\,(\ref{seq15}).

To prove that the effects of $\hat{H}(t)$ on $\ket{\psi(t)}$ is to make $\ket{\psi(t)}$ rotate around $\vec{v}_H$, we note that by Eq.\,(\ref{eq2}),
\begin{equation}
	\hat{H}^2(t)=\ket{\p_t\psi(t)}\bra{\p_t\psi(t)}+\omega^2(t)\ket{\psi(t)}\bra{\psi(t)},
\end{equation}
\begin{equation}
	\hat{H}^3(t)=\omega^2(t) \hat{H}(t),
\end{equation}
where
\begin{equation}
	\omega^2(t)=\braket{\p_t \psi(t)}{\p_t \psi(t)},
\end{equation}
hence there exists an recursing pattern for the power of $\hat{H}(t)$. By making use of this pattern, and assuming for the moment that $\hat{H}(t)$ to be a constant, we have
\begin{equation}\label{seq11}
	\begin{aligned}
U(T)&=\exp(-i H(t)T)=1-iH(t)T+\f{1}{2!}(-i)^2H^2(t)T^2+\f{1}{3!}(-i)^3H^3(t)T^3+\dots\\
&=1-iH(t)T-\f{1}{2!}H^2(t)T^2-\f{1}{3!}i\omega^2(t)H(t)T^3+\f{1}{4!}\omega^2(t)H^2(t)T^4-\f{1}{5!}i\omega^4(t)H(t)T^5+\dots.
\end{aligned}
\end{equation}
such that 
\begin{equation}
	\begin{aligned}
\ket{\psi(t+T)}&=U(T)\ket{\psi(t)}\\&=\ket{\psi(t)}+\ket{\p_t\psi(t)}T-\f{1}{2!}\omega^2(t)\ket{\psi(t)}T^2+\f{1}{3!}\omega^2(t)\ket{\p_t\psi(t)}T^3+\f{1}{4!}\omega^4\ket{\psi(t)}T^4+\cdots\\
&=\alpha \ket{\psi(t)}+\beta \ket{\p_t \psi(t)}
\end{aligned}
\end{equation}
where $\alpha$ and $\beta$ are some real numbers that we do not need to solve explicitly. Since $\bra{\psi(t+T)} \hat{H}(t)\ket{\psi(t+T)}=0$, it indicates $\vec{v}_H$ is orthogonal to $\vec{v}_{\rho_{t+T}}$.

Of course, $\hat{H}(t)$ could be time-dependent, so that the direct application of Eq.\,(\ref{seq11}) may be inappropriate. In that case, we can let $T$ be an infinitesimally small number $\delta t$ so that the above argument is valid during each infinitesimally small time interval $[t, t+\delta t], [t+\delta t, t+2\delta t],\cdots$. The scenario then becomes that $\vec{v}_\rho$ is always orthogonal to and rotates around $\vec{v}_H$, and meanwhile $\vec{v}_H$ itself is changing its pointing direction in the Bloch sphere.
\\\\
\begin{center}
\textbf{\large Supplemental Materials V}
\end{center}
Let the initial and final state of a qubit be denoted by
$\hat{\rho}(0)=\f{1}{2}(\mathbbm{1}+x_\text{I}\hat{\sigma}_x+y_\text{I}\hat{\sigma}_y+z_\text{I}\hat{\sigma}_z)$ and $\rho(T)=\f{1}{2}(\mathbbm{1}+x_\text{F}\hat{\sigma}_x+y_\text{F}\hat{\sigma}_y+z_\text{F}\hat{\sigma}_z)$, respectively. Consider the case where the Hamiltonian of the qubit is given by $\hat{H}=\f{\epsilon}{2} \hat{\sigma}_z-\f{\Delta}{2}\hat{\sigma}_x$ and the measuring operator $\hat{\mathcal{M}}_{\delta t}$ is given by $\hat{\mathcal{M}}_{\delta t}=(\delta t / 2 \pi \tau)^{1 / 4} \exp \left[-\frac{\delta t}{4 \tau}\left(r_t\mathbbm{1}-\hat{\sigma}_{z}\right)^{2}\right]$. The qubit being continuously measured will then evolve under both the effect from its own Hamiltonian and measuring processes, such that
\begin{equation}
	\hat{\rho}(t+\delta t)=\frac{\hat{\mathcal{U}}_{\delta t} \hat{\rho}(t) \hat{\mathcal{U}}_{\delta t}^{\dagger}}{\operatorname{Tr}\left[\hat{\mathcal{U}}_{\delta t} \hat{\rho}(t) \hat{\mathcal{U}}_{\delta t}^{\dagger}\right]} \quad \text{ where } \quad \hat{\mathcal{U}}_{\delta t} \equiv e^{-\frac{i}{\hbar} \hat{H} \delta t} \hat{\mathcal{M}}_{\delta t},
\end{equation}
from which one can obtain a master equation
\begin{equation}
	\partial_{t} \hat{\rho}=-\frac{i}{\hbar}[\hat{H}, \hat{\rho}]+\frac{r_t}{2 \tau}\left\{\hat{\sigma}_{z}, \hat{\rho}\right\}-\frac{r_t}{\tau}\left\langle\hat{\sigma}_{z}\right\rangle \hat{\rho}.
\end{equation}
It is shown in \cite{PhysRevA.88.042110} that the most probable path of the qubit subject to the above master equation is given by
\begin{equation}
	\begin{aligned} \bar{x}(t) &=\frac{x_{I} \cos \epsilon t-y_{I} \sin \epsilon t}{\cosh \bar{r} t / \tau+z_{I} \sinh \bar{r} t / \tau} \\ \bar{y}(t) &=\frac{y_{I} \cos \epsilon t+x_{I} \sin \epsilon t}{\cosh \bar{r} t / \tau+z_{I} \sinh \bar{r} t / \tau} \\ \bar{z}(t) &=\frac{z_{I} \cosh \bar{r} t / \tau+\sinh \bar{r} t / \tau}{\cosh \bar{r} t / \tau+z_{I} \sinh \bar{r} t / \tau}, \end{aligned}
\end{equation}
\begin{equation}
	\text{ where }\quad  \bar{r}=\frac{\tau}{T} \tanh ^{-1}\left(\frac{z_{I}-z_{F}}{z_{I} z_{F}-1}\right) \quad \text{ if }\quad \Delta=0.
\end{equation}
To calculate the effective Hamiltonian $\hat{H}_\text{eff}(t)$ responsible for the most probable path, let us consider the case where $\Delta=0$ and  $x_\text{I}=1, y_\text{I}=0,$ and $z_\text{I}=0$, which corresponds to the pure initial state $\ket{\psi(0)}=\f{1}{\sqrt{2}}(\ket{0}+\ket{1})$. The density matrix at time $t$ then is
\begin{equation}
	\rho(t)=\f{1}{2}(\mathbbm{1}+\bar{x}(t)\hat{\sigma}_x+\bar{y}(t)\hat{\sigma}_y+\bar{z}(t)\hat{\sigma}_z).
\end{equation}
Let $\ket{\varphi_i(t)}$ be the normalized instantaneous eigenvector of $\rho(t)$, from which we can obtained $\ketz{\tilde{\varphi_i}(t)}$ defined by $\ketz{\tilde{\varphi_i}(t)}\equiv e^{i \phi(t)} \ket{\varphi_i(t)}$ where $\phi(t)\equiv \int -i \braket{\p_t \varphi_i(t)}{\varphi_i(t)}\,dt$. The effective Hamiltonian $\hat{H}_\text{eff}(t)$ driving $\rho(t)$ then can be given by \cite{hu2022quantum}
\begin{equation}
	\begin{aligned}
\hat{H}_\text{eff}(t)&=\f{i}{2}\sum_{i=1}^2(\ketz{\p_t\tilde{\varphi_i}(t)}\braz{\tilde{\varphi_i}(t)}-\ketz{\tilde{\varphi_i}(t)}\braz{\p_t \tilde{\varphi_i}(t)})\\
&=\begin{pmatrix}
	\f{\epsilon}{2}\sech^2\alpha(t,z_\text{F})) & \f{i e^{-i t \epsilon}}{2T}(\tanh^{-1}{z_F}+iT\epsilon \tanh\alpha(t,z_\text{F}))\sech(\alpha(t,z_\text{F}))\\
	\f{-i e^{i t \epsilon}}{2T}(\tanh^{-1}{z_F}-iT\epsilon \tanh(\alpha(t,z_\text{F}))\sech(\alpha(t,z_\text{F})) & -\f{\epsilon}{2}\sech^2(\alpha(t,z_\text{F}))
\end{pmatrix}
\end{aligned}
\end{equation}
where \begin{equation}
	\alpha(t,z_\text{F})=\f{t}{T} \tanh^{-1}{z_\text{F}}.
\end{equation}
It is worth noting that the instantaneous energy variance of $\hat{H}_\text{eff}(t)$ is given by
\begin{equation}
	[\Delta H_\text{eff}(t)]^2=\operatorname{Tr}(\rho(t)\hat{H}^2_\text{eff}(t))-\operatorname{Tr}(\rho(t)\hat{H}_\text{eff}(t))^2=\frac{\left(T^{2} \epsilon^{2}+[\tanh ^{-1}z_\text{F}]^{2}\right) \operatorname{sech}^{2}\left(\alpha(t,z_\text{F})\right)}{4 T^{2}},
\end{equation}
which indicates the smaller the $T$ and the larger the $z_\text{F}$ are, the larger the $[\Delta H_\text{eff}(t)]^2$ is.
\\\\
\begin{center}
\textbf{\large Supplemental Materials VI}
\end{center}
In the following, we will show more details in calculations of Example III in the main text.

Let the initial wave function $\psi(x,0)$ at time $t=0$ be peaked at $x_0$,
\begin{equation}
	\psi(x,0)=\left(\frac{m \omega}{\pi \hbar}\right)^{1 / 4} \exp \left(-\frac{m \omega}{2 \hbar} (x-x_0)^{2}\right),
\end{equation}
such that we can obtain $\ket{\psi(0)}$ by $\ket{\psi(0)}=\int \psi(x,0)\ket{x}dx$.
Let the measurement operator $\hat{\mathcal{M}}_{\delta t}$ be 
\begin{equation}
	\hat{\mathcal{M}}_{\delta t}=\int (\delta t / 2 \pi \tau)^{1 / 4} \exp \left[-\frac{\delta t}{4 \tau}\left(r_t-x\right)^{2}\right] \ket{x}\bra{x}dx.
\end{equation}
We need to calculate
\begin{equation}
	\ketz{\p_t {\psi}(0)}=\lim_{\delta t \to 0} \f{\ket{\psi(\delta t)}-\ket{\psi(0)}}{\delta t}=\lim_{\delta t \to 0}\f{\left(\f{\hat{\mathcal{M}}_{\delta t}}{\sqrt{\int\bra{\psi(0)}\hat{\mathcal{M}}_{\delta t}^\dagger \hat{\mathcal{M}}_{\delta t} \ket{\psi(0)}}}-1\right)}{\delta t}  \ket{\psi(0)}
\end{equation}
Since
\begin{equation}
	\lim_{\delta t \to 0}\left(\f{\hat{\mathcal{M}}_{\delta t}}{\sqrt{\bra{\psi(0)}\hat{\mathcal{M}}_{\delta t}^\dagger \hat{\mathcal{M}}_{\delta t} \ket{\psi(0)}}}-1\right)/\delta t=\int \f{\hbar+2 m \omega(x-x_0)(2r_t-x-x_0)}{8m\omega \tau}\ket{x}\bra{x}dx+\mathcal{O}(\delta t),
\end{equation}
by Eq.\,(\ref{eq2})(\ref{eq3}) we have
\begin{equation}
	\ket{\p_t \psi(0)}=\int \underbrace{\f{\hbar+2 m \omega(x-x_0)(2r_t-x-x_0)}{8m\omega \tau}\left(\frac{m \omega}{\pi \hbar}\right)^{1 / 4} \exp \left(-\frac{m \omega}{2 \hbar} (x-x_0)^{2}\right)}_{\dot{\psi}(x,0)} \ket{x}dx.
\end{equation}
Since $\braket{\psi{(0)}}{\p_t \psi(0)}=0$, we have again $\ketz{\tilde{\psi}(0)}=\ket{\psi(0)}$.
Therefore, according to Eq.~(2), the measuring Hamiltonian at time $t=0$ is given by

\begin{equation}
\begin{aligned}
\hat{H}(0)&=i\hbar(\ketz{\p_t\tilde{\psi}(0)}\braz{\tilde{\psi}(0)}-\ketz{\tilde{\psi}(0)}\braz{\p_t \tilde{\psi}(0)})\\
&=i \hbar\int (\dot{\psi}(x,0)\psi(x^\prime,0)-\psi(x,0)\dot{\psi}(x^\prime,0))\ket{x}\bra{x^\prime}dxdx^\prime\\
&=i\hbar\int \f{\eta_{x x^\prime}}{4\tau}\sqrt{\f{m \omega}{\pi \hbar}}\exp{\left( -\f{m \omega ((x-x_0)^2+{(x^\prime}-x_0)^2)}{2\hbar}\right)} \ket{x}\bra{x^\prime}dxdx^\prime
\end{aligned}\end{equation}
where $\eta_{xx^\prime} \equiv (x-x^\prime)(2r_t-x-x^\prime).$
We can then calculate instantaneous energy uncertainty,
\begin{equation}
	\Delta H(0)=\|\hbar\ket{\p_t \psi(0)}\| =\sqrt{\int |\hbar \dot{\psi}(x,0)|^2dx}=\sqrt{\frac{\hbar \left(\hbar+4 m \omega (r_t-x_0)^{2} \right)}{32 m^{2} \omega^{2}\tau^{2} }}\hbar
\end{equation}
and the wave function at time $t=\delta t$,
\begin{equation}
	\ket{\psi(\delta t)}=\int \f{1}{\Delta X(\delta t)\sqrt{2\pi}} \exp{\left[-\f{1}{2}\left(\f{x-\langle X(\delta t) \rangle}{\Delta X(\delta t)}\right)^2\right]}\ket{x}dx,
\end{equation}
which is still a Gaussian, with mean position
\begin{equation}
	\langle X(\delta t) \rangle =\f{r_t \hbar \,\delta t+2 m \tau \omega x_0}{\hbar \,\delta t+2m \tau \omega}
	\end{equation}
and position uncertainty,
\begin{equation}
\Delta X(\delta t)^2=\langle X^2(\delta t) \rangle-\langle X(\delta t) \rangle^2 =\f{\hbar \tau}{\hbar\, \delta t+2 m \tau \omega}.
\end{equation} 
On the other hand, since
\begin{equation}
	\Delta X(0)^2=\f{\hbar}{2 m \omega},
\end{equation}
hence
\begin{equation}
	\delta \Delta X(0)^2=\f{\hbar \tau}{\hbar \delta t+2 m \tau \omega}-\f{\hbar}{2 m \omega}
\end{equation}
such that
\begin{equation}
	\f{\p ([\Delta X(0)]^2)}{\p t} = \lim_{\delta t \to 0}\f{\delta ([\Delta X(0)]^2)}{\delta t}=-\f{\hbar^2}{4m^2 \omega^2 \tau}=- [\Delta X(0)^2]^2 \f{1}{\tau}
\end{equation}
which becomes an ordinary differential equation. It can be solved as
\begin{equation}
	\Delta X(t)^2=\f{\Delta X(0)^2 \tau}{\Delta X(0)^2 t+\tau}
\end{equation}
The above calculation implies that if we apply sequential $\hat{\mathcal{M}}_{\delta t}$ on $\ket{\psi(0)}$, the wave function at any later time will still be a Gaussian, and only its mean and variance changes, i.e.
\begin{equation}
\ket{\psi(t)} \simeq \int \f{1}{\Delta X_t \sqrt{2 \pi}}\exp{\left(- \f{(x-\lra{x_t}))^2}{2 \Delta X_t^2}\right)} \ket{x}dx.
\end{equation}
Since we can relate $\omega$ with $\Delta X(t)$ by
\begin{equation}
	\omega=\f{\hbar}{2 m \Delta X(t)^2}.
\end{equation}
Plugging in the expression to $\Delta H(0)$, we obtain $\Delta H(t)$
\begin{equation}
	\Delta H(t) = \frac{\sqrt{2 (r_t-x_0)^{2} \Delta X_t^{2}+\Delta X_t^{4}}}{2 \sqrt{2} \tau}\hbar,
\end{equation}
At the strong measurement limit $\tau \to 0$, we have
\begin{equation}
	\lim_{\tau \to 0}\Delta H (t)=\f{|r_t-x_t|}{2\sqrt{\tau t}} \hbar+\mathcal{O}(\tau).
\end{equation}

\end{document}